\documentclass[aps,reprint,showkeys,amsmath,amssymb,groupedaddress]{revtex4-1}
\pdfoutput=1
\usepackage{xcolor, siunitx, microtype}
\usepackage[pdfauthor={Pedro Mediano},
            pdftitle={Integrated Information and Metastability in Systems of Coupled Oscillators},
            pdfkeywords={Synchronisation, chimera states, metastability, integrated information},
          ]{hyperref}

\DeclareMathOperator{\var}{var}

\begin{document}


\title{Integrated Information and Metastability in Systems of Coupled Oscillators}

\author{Pedro A.M. Mediano}
\email{pmediano@imperial.ac.uk}
\author{Juan Carlos Farah}
\author{Murray Shanahan}
\affiliation{Department of Computing, Imperial College London}
\date{\today}

\begin{abstract}

It has been shown that sets of oscillators in a modular network
can exhibit a rich variety of metastable chimera states, in which
synchronisation and desynchronisation coexist. Independently, under the guise
of integrated information theory, researchers have attempted to quantify the
extent to which a complex dynamical system presents a balance of integrated and
segregated activity. In this paper we bring these two areas of research
together by showing that the system of oscillators in question exhibits a
critical peak of integrated information that coincides with peaks in other
measures such as metastability and coalition entropy.

\end{abstract}


\maketitle

\section{Introduction}

Systems of coupled oscillators are ubiquitous both in nature and in the
human-engineered environment, making them of considerable scientific interest
\cite{Pikovsky2001}. A variety of mathematical models of such systems have been
devised and their synchronisation properties have been the subject of much
study. Typical studies of this sort, such as the classic work of Kuramoto
\cite{Kuramoto1984}, examine the conditions under which the system converges on
a stable state of either full synchronisation or desynchronisation, perhaps
identifying an order parameter that determines a critical phase transition from
one state to the other. The collection of known attractors of such systems was
enlarged with the discovery of so-called chimera states, in which the system of
oscillators partitions into two stable subsets, one of which is fully
synchronised while the other remains permanently desynchronised
\cite{Panaggio2015}.

Although systems of coupled oscillators that converge on stable states are both
mathematically interesting and scientifically relevant, they are by no means
representative of all real-world synchronisation phenomena. For example, the
brain exhibits synchronous rhythmic activity on multiple spatial and temporal
scales, but never settles into a stable state. Although it enters chimera-like
states of high partial synchronisation, these are only temporary. A system of
coupled oscillators that continually moves from one highly synchronised state
to another under its own intrinsic dynamics is said to be metastable. In
\cite{Shanahan2010}, it was shown that a modular network of phase-lagged
Kuramoto oscillators will exhibit metastable chimera states under certain
conditions. Variants of this model have since been used to replicate the
statistics of the brain under a variety of conditions, including the resting
state \cite{Cabral2011}, cognitive control \cite{Hellyer2015}, and anaesthesia
\cite{Schartner2015}.

In a separate line of enquiry, a number of researchers have attempted to pin
down the notion of dynamical complexity. A system is said to have high
dynamical complexity if it exhibits a balance of integrated and segregated
activity, where a system's activity is integrated to the extent that its parts
influence each other and segregated to the extent that its parts act
independently \cite{Shanahan2008}. Prominent among these attempts is the
Integrated Information Theory (IIT), originally proposed by Balduzzi and Tononi
\cite{Balduzzi2008a}, but extended by multiple authors ever since
\cite{Barrett2011, Griffith2014, Tegmark2016}.

In the present paper, we connect these two lines of enquiry by demonstrating
that modular networks of coupled oscillators of the sort described in
\cite{Shanahan2010} not only exhibit metastable chimera states, but also have
high dynamical complexity. Moreover, we show that measures of both phenomena
peak in the narrow critical region of the parameter space wherein the system is
poised between order and disorder, and IIT offers a rich picture of dynamical
complexity in this critical regime. To our knowledge, this is the first
description of a dynamical system in which the three major complexity
indicators of criticality, metastability, and integrated information all
appear.

\section{Methods}

We examine a system of coupled Kuramoto oscillators, extensively used to study
non-linear dynamics and synchronisation processes \cite{Abrams2008}. We build
upon the work of \cite{Shanahan2010} with a community-structured network of
oscillators. The network is composed of 8 communities of 32 oscillators each,
with every oscillator being coupled to all other oscillators in its community
with probability 1 and to each oscillator in the rest of the network with
probability $1/32$. The state of each oscillator $i$ is captured by its phase
$\theta_i$, the evolution of which is governed by the equation
\begin{align}
  \frac{d \theta_i}{d t} = \omega + \frac{1}{\kappa + 1} \displaystyle\sum_j K_{ij} \sin \left( \theta_j - \theta_i - \alpha \right) ~ ,
  \label{eq:kur}
\end{align}
\noindent where $\omega$ is the \emph{natural frequency} of each oscillator,
$\kappa$ is the average degree of the network, $K$ is the \emph{connectivity
matrix} and $\alpha$ is a global \emph{phase lag}. We set $\omega = 1$ and
$\kappa=63$. To reflect the community structure, the coupling between two
oscillators $i,j$  is $K_{ij} = 0.6$ if they are in the same community or
$K_{ij} = 0.4$ otherwise.  We tune the system by modifying the value of the
phase lag, parametrised by $\beta = \pi/2 - \alpha$. We note that the system is
fully deterministic, i.e.  there is no noise injected in the dynamical
equations.

\subsection{Metastability}

In this section we review the basic concepts behind metastability and how it
is quantified, following a similar description to that of \cite{Shanahan2010}.

The building block of the dynamical quantities we study in this article is the
\emph{instantaneous synchronisation} $R$, that quantifies the dispersion in
$\theta$-space of a given set of oscillators. In general, we denote as
$R_c(t)$ the instantaneous synchronisation of a community $c$ of oscillators at
time $t$, given by
\begin{equation}
  R_c(t) = | \langle e^{i \theta_j(t) } \rangle_{j \in c} | ~ .
  \label{eq:sync}
\end{equation}
To quantify metastability, we use the \emph{metastability index} $\lambda$,
which is defined as the average temporal variance of the synchrony of each
community $c$, i.e. 
\begin{subequations}
\begin{gather}
  \lambda_c = \var_t R_c(t) \label{eq:lambdac} \\
  \lambda = \langle \lambda_c \rangle_c ~ .
\end{gather}
\label{eq:lambda}%
\end{subequations}
\noindent Last, we also define \emph{global synchrony} $\xi$ as the average
across time and space of instantaneous synchrony,
\begin{equation}
  \xi = \big| \big\langle R_c(t) \big\rangle_{t,c} \big| ~ .
\end{equation}

According to Eq.~\eqref{eq:sync}, $R_c$ (and therefore $\xi$) is bounded in the
$[0,1]$ interval. $R_c(t)$ will be 1 if all oscillators in $c$ have the same
phase at time $t$, and will be 0 if they are maximally spread across the unit
circle. This $[0,1]$ bound on $R$ allows us to place an upper bound on
$\lambda$ -- assuming a unimodal synchrony distribution, the maximum possible
value of $\lambda$ is $\lambda_{max} = 1/9$.

As defined in Eq.~\eqref{eq:lambdac}, $\lambda_c$ represents the size of the
fluctuations in the internal synchrony of a community. A system that is either
hypersynchronised or completely desynchronised will have a very small
$\lambda_c$, whereas one whose elements fluctuate in and out of synchrony will
have a high $\lambda_c$. In other words, a system of oscillators exhibits
metastability if its elements remain in the vicinity of a synchronised state
without falling into such a state permanently.

\subsection{Integrated Information}

Although other information-theoretic quantities have also been linked to
complexity in a neuroscience context \cite{King2013,Schartner2015}, we take
integrated information $\Phi$ as the main informational measure of complexity
in our study \cite{Balduzzi2008a}. There are more modern accounts of the theory
\cite{Oizumi2014,Tononi2012}, but the latest versions have not been as
thoroughly studied and are not amenable to easy estimation from time series
data. For these reasons, we focus on the methods in \cite{Barrett2011} to
empirically estimate $\Phi$ from an observed time series.

The building block of integrated information is \emph{effective information},
$\varphi$. Effective information quantifies how much better a system $X$ is at
predicting its own future (or decoding its own past) after a time $\tau$ when
it is considered as a whole compared to when it is considered as the sum of two
subsystems $M^{\{1,2\}}$. We refer to $\tau$ as the \emph{integration
timescale}. In other words, $\varphi$ evaluates how much information is
generated by the system but not by the two subsystems alone. For a specific
bipartition $\mathcal{B} = \{M^1,M^2\}$, the effective information of the
system \emph{beyond} $\mathcal{B}$ is
\begin{align}
  \varphi[X; \tau, \mathcal{B}] = I(X_{t-\tau}, X_t) - \displaystyle\sum_{k=1}^2 I(M_{t-\tau}^k,
  M_t^k) ~ .
  \label{eq:ei}
\end{align}
The main idea behind the computation of $\Phi$ is to exhaustively search all
possible partitions of the system and calculate the effective information of
each of them. Among those we select the partition with lowest $\varphi$ (under
some considerations, see below), termed the \emph{Minimum Information
Bipartition} (MIB). Then, the integrated information of the system is the
effective information beyond its MIB. Given the above expression for $\varphi$,
$\Phi$ is defined as
\vspace{2pt}
\begin{subequations}
\begin{gather}
\Phi[X,\tau] = \varphi[X; \tau, \mathcal{B}^{\mathrm{MIB}}] \\
\mathcal{B}^{\mathrm{MIB}} = \arg_{\mathcal{B}}\min \frac{\varphi[X; \tau, \mathcal{B}]}{K(\mathcal{B})} \\
  K(\mathcal{B}) = \min \left\{ H(M^1), H(M^2) \right\} ,
\end{gather}
\label{eq:phi}%
\end{subequations}
\noindent where $K$ is a normalisation factor to avoid biasing $\Phi$ to excessively
unbalanced bipartitions. Defined this way, $\Phi$ can be understood as the
minimum information loss incurred by splitting the system into two subsystems.
It quantifies the collective emergent behaviour that is present in the whole
system but not in any bipartition.

\section{Results}
\label{sec:res}

We ran 1500 simulations with values of $\beta$ distributed uniformly at random
in the range $[0, 2\pi)$ using RK4 with a stepsize of 0.05 for numerical
integration. Each simulation was run for \num{5e6} timesteps, of which the
first \num{e4} are discarded to avoid effects from transient states. All
information-theoretic measures are reported in bits.

\begin{figure}[ht]
\centering
\includegraphics{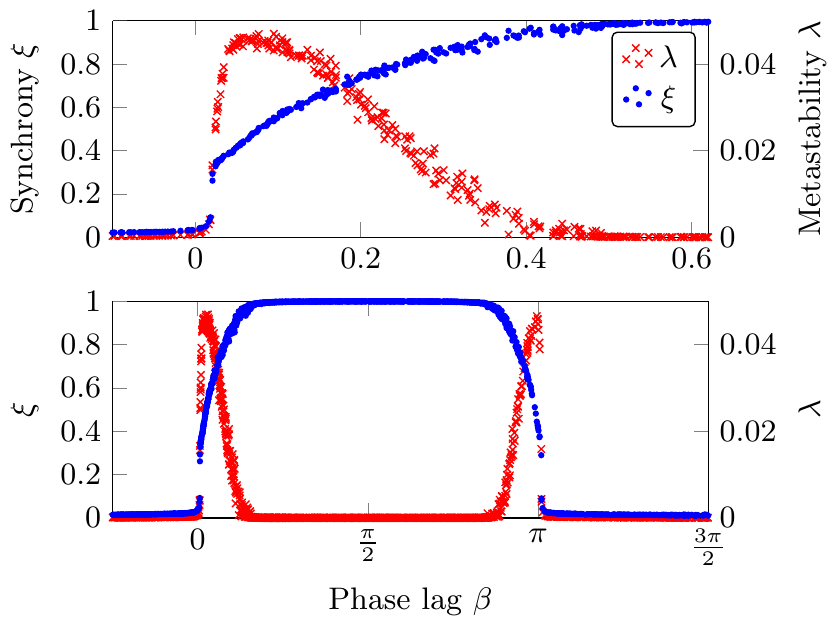}
\caption{Global synchrony and metastability for different phase lags $\beta$
for the whole $[0, 2\pi)$ range (bottom) and around the critical transition
region (top). Rapid increase of metastability marks the onset of the phase
transition. Note the different $\beta$ ranges in both plots.}
\label{fig:sync}
\end{figure}

We first study the system from a purely dynamical perspective, following the
analysis in \cite{Shanahan2010}. Global synchrony and metastability are shown
in Fig.~\ref{fig:sync}. The first characteristic we observe is that there
are two well differentiated dynamical regimes -- one of hypersynchronisation
and one of complete desynchronisation, with strong metastability appearing in
the narrow transition bands between one and the other.

It is in this transition region where the oscillators operate in a critical
regime poised between order and disorder and complex phenomena appear. As the
system moves from desynchronisation to full synchronisation there is a sharp
increase in metastability, followed by a smoother decrease as the system becomes
hypersynchronised. In the region $0 < \beta < \pi/8$, the system remains in a
complex equilibrium between an ordered and a disordered phase.

\subsection{Information-theoretic analysis}

One feature of $\Phi$ (and of any other information-theoretic measure),
compared to $\lambda$, is that it does not need to be calculated directly from
the state of the system. According to its definition, $\Phi$ is
\emph{substrate-agnostic}, meaning that the relevant quantity for the
calculation of $\Phi$ is not the physical state of the system, but some
\emph{informational state} -- the configuration of the system that we consider
to contain information. Therefore, we must define an \emph{informational state
mapping}, that extracts the information-bearing symbols from the physical state
of the system.

Although calculating $\Phi$ on the real-valued phases is possible, for
simplicity we choose the \emph{coalition configuration} of the system as the
informational state, defined as the set of communities that are highly
internally synchronised. To calculate the coalition configuration at time $t$
we calculate $R_c(t)$ of each community and threshold it, such that
\begin{equation*}
  X_t^c = \begin{cases} 1 & \text{if } R_c(t) > \gamma\\ 0 &
    \text{otherwise.} \end{cases}
\end{equation*}

We refer to $\gamma$ as the \emph{coalition threshold}. After calculating the
coalitions, the history of the system is reduced to a time series with 8 binary
variables. Having a multivariate discrete time series, it is now tractable to
compute $\Phi$. By default, we use $\gamma = 0.8$ in all our analyses shown
here.

\begin{figure}[ht]
  \centering
  \includegraphics{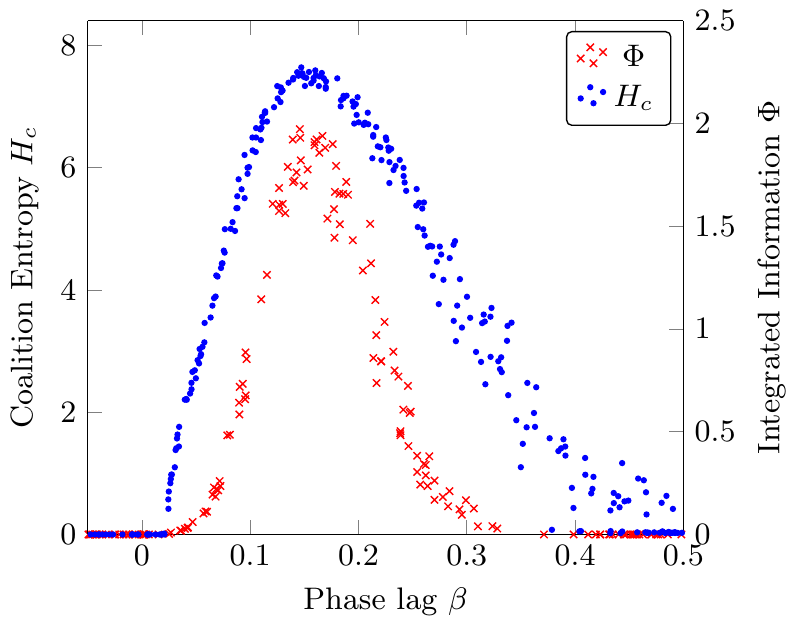}
  \caption{Integrated information $\Phi$ and coalition entropy $H_c$ in the
  phase transition. Within the broad region between order and disorder in which
  $H_c$ rises there is a narrower band in which complex spatiotemporal patterns
  generate high $\Phi$.}
\label{fig:phi}
\end{figure}

As depicted in Fig.~\ref{fig:phi}, $\Phi$ shows a similar behaviour to
$\lambda$ -- it peaks in the transition regions and shrinks in the fully
ordered and the fully disordered regimes. We also compare $\Phi$ with arguably
the simplest information-theoretic measure -- entropy $H$. The entropy of the
state of the network calculated on the coalitions $X_t$ forms the
\emph{coalition entropy} $H_c$.

Both $\Phi$ and $H_c$ peak precisely at the same point. Although both measures
depend on the chosen coalition threshold $\gamma$, the results are
qualitatively the same for a wide range of thresholds.

Although it peaks in the same region as $\lambda$ and $H_c$, we note that
$\Phi$ reveals new properties of the system by virtue of incorporating temporal
information in its definition. That is, to have a high $\Phi$ a system must
exhibit complex spatial \emph{and} temporal patterns.  We can verify this by
performing a random time-shuffle on the time series. This shuffling leaves
$\lambda$ and $H_c$ unaltered, as they don't explicitly depend on time
correlations, but has a high impact on $\Phi$, which shrinks to zero. This
indicates that $\Phi$ is sensitive to properties of the system that are not
reflected by other measures.

\begin{figure}[ht]
  \centering
  \includegraphics{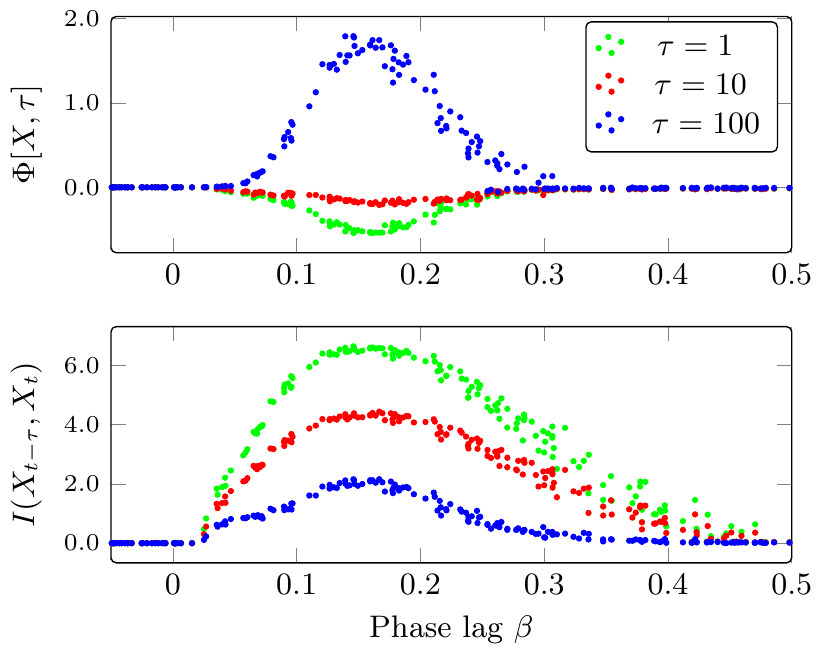}
  \caption{Integrated information $\Phi$ and time-delayed mutual information
    $I(X_{t - \tau}, X_t)$ for several timescales $\tau$. See text for
    details.}
  \label{fig:phiManyTau}
\end{figure}

Furthermore, $\Phi$ can be used to investigate the behaviour of the system at
multiple timescales.  Figure \ref{fig:phiManyTau} shows the behaviour of $\Phi$
for several values of $\tau$, and compares it with standard time-delayed mutual
information (TDMI) $I(X_{t - \tau}, X_t)$.

The first thing we note is that $\Phi$ and TDMI have opposite trends with
$\tau$. TDMI decreases for longer timescales while $\Phi$ increases. At short
timescales the system is highly predictable -- thus the high TDMI -- but this
short-term evolution does not involve any system-wide interaction -- thus the
low $\Phi$. Furthermore, at these timescales $\Phi$ is negative, which can be
interpreted as an indication of \emph{redundancy} \cite{Barrett2014} in the
evolution of the system: the parts share some information, such that they
separately contain more information about their past than the whole system
about its past. For larger $\tau$ TDMI decreases, as the evolution of the
system becomes harder to track using the coalition configuration.
Simultaneously, $\Phi$ becomes higher, indicating that the remaining TDMI has a
stronger integrated component that is not accounted for by the TDMI of the
partitions of the system. Overall, we see a clear trend of TDMI diminishing at
longer timescales but becoming progressively more integrated in nature.


It might seem counterintuitive to the reader that both TDMI and $\Phi$ converge
to a non-zero value for arbitrarily high $\tau$. However, recall that the
system is deterministic, so there is no reason to expect TDMI to vanish even in
the $\tau\rightarrow\infty$ limit. Perfect knowledge of all of the oscillators'
phases at any given time step is enough to reconstruct the whole history of the
system. The reason why we see a decreasing TDMI is because the informational
state we chose (the coalition configuration) is not descriptive enough to
capture the evolution of the system perfectly. That is, this is not a result of
stochasticity, but of degeneracy.

Put another way, the TDMI of the whole system remains non-zero in the
$\tau\rightarrow\infty$ limit because we are dealing with a causally closed
system. Again, if we considered the totality of the system's state (i.e. the
phases $\theta_i$) then TDMI would be maximal and constant for any $\tau$. In
contrast, when considering the TDMI of any partition $M$ we start dealing with
a causally open system, since one partition is affected by the other. This
effectively introduces stochasticity in our observations $M_t$, which does
cause TDMI of the partition to vanish when $\tau\rightarrow\infty$. This
explains that the TDMI of the whole system converges to a non-zero value for
large $\tau$ while the TDMI of any partition fades to zero, leaving a positive
$\Phi$.


Finally, it is interesting to combine the insights from the dynamical and
information-theoretic analyses. Inspecting Figs.~\ref{fig:sync} and
\ref{fig:phi} we see that the peak in $\Phi$ is much narrower than the peaks in
$\lambda$ and $H_c$. While some values of $\beta$ do give rise to non-trivial
dynamics, it is only at the centre of the critical region that these dynamics
give rise to integration. A certain degree of internal variability is necessary
to establish integrated information, but not all configurations with high
internal variability lead to a high $\Phi$. This means that $\Phi$ is sensitive
to more complicated dynamic patterns than the other measures considered, and is
in that sense more discriminating.

We note that $\lambda$ is a community-local quantity -- that is, the
calculation of $\lambda_c$ for each community is independent of the rest.
Conversely, $\Phi$ relies exclusively on the irreducible interaction between
communities. These two quantities are nevertheless intrinsically related,
insofar as internal variability enables the system to visit a larger repertoire
of states in which system-wide interaction can take place.

\subsection{Robustness of $\Phi$ against measurement noise}

\begin{figure}[ht]
  \centering
  \includegraphics{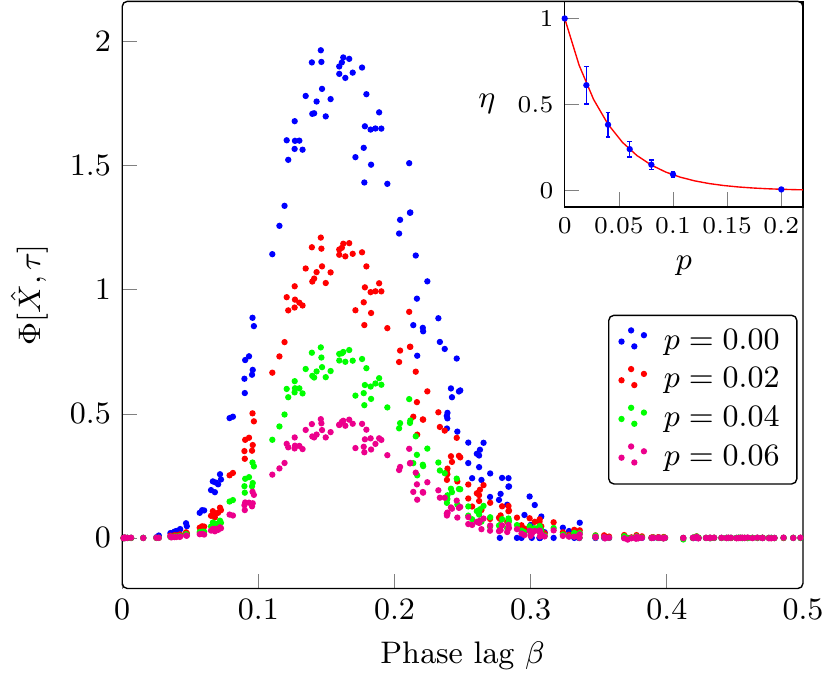}
  \caption{Integrated information $\Phi$ for different levels of measurement
    noise $p$. Inset: (\textcolor{blue}{blue}) Mean and variance of the ratio
    $\eta$ between $\Phi$ of the corrupted and the original time series.
    (\textcolor{red}{red}) Exponential fit $\eta = \exp(-p/\ell)$, with $\ell
    \approx 0.04$.}
  \label{fig:phiNoise}
\end{figure}

We will now consider the impact of measurement noise on $\Phi$, wherein the
system runs unchanged but our recording of it is imperfect. For this experiment
we run the (deterministic) simulation as presented in the previous section and
take the binary time series of coalition configurations. We then emulate the
effect of uncorrelated measurement noise by flipping each bit in the time
series with probability $p$, yielding the corrupted time series $\hat{X}$.
Finally we recalculate $\Phi$ on the corrupted time series, and show the
results in Fig.~\ref{fig:phiNoise}.  To quantify how fast $\Phi$ changes we
calculate the ratio between the corrupted and the original time series,
\begin{equation}
  \eta = \frac{\Phi[\hat{X}, \tau]}{\Phi[X, \tau]} ~ .
  \label{eq:eta}
\end{equation}

In order to avoid instabilities as $\Phi[X, \tau]$ gets close to zero, we
calculate $\eta$ only in the region within \SI{0.5}{\radian} of the centre of the peak,
where $\Phi[X, \tau]$ is large.  The inset of Fig.~\ref{fig:phiNoise} shows the
mean and standard deviation of $\eta$ at different noise levels $p$.

We find that $\Phi$ monotonically decays with $p$, reflecting the gradual loss
of the precise spatiotemporal patterns characteristic of the system. The
distortion has a greater effect on time series with greater $\Phi$, but preserves
the dominant peak in $\beta \approx 0.15$. The inset shows that both the mean
and variance of $\eta$ decay as a clean exponential with $p$. $\Phi$ is highly
sensitive to noise and undergoes a rapid decline, as a measurement noise of 5\%
wipes out 70\% of the perceived integrated information of the system.

\section{Conclusion}

We have presented a community-structured network of Kuramoto oscillators and
discussed their collective behaviour in terms of metastability
\cite{Shanahan2010} and integrated information \cite{Balduzzi2008a}. We showed
that the system undergoes a phase transition whose critical region presents a
sharp, clear peak of integrated information $\Phi$ that coincides with a strong
increase in metastability. To our knowledge, this is the first description of a
dynamical system in which the three major complexity indicators of criticality,
high metastability, and high integrated information all appear. The resulting
confluence of two major research directions in complexity science suggests that
this is a system that merits further study.

In the context of the present model, the high internal variability of the
system's components enables system-wide interaction, which in turn leads to
high $\Phi$. As we have seen, the system presents a region of high
metastability, but notably it is only within an even narrower band that we find
strong integrated information. Moreover, as we have also seen, shuffling the
time series data preserves the peak of metastability, despite the fact that the
result is meaningless. By contrast, $\Phi$ only peaks when the relevant
temporal structure is present in the data. In this way we provide evidence that
complex dynamics -- as quantified by the metastability index $\lambda$ -- are a
necessary but not sufficient condition for complex information processing -- as
quantified by integrated information $\Phi$.
 
Dynamical and information-theoretic measures provide different lenses through
which we can understand a system, and offer complementary views of its
behaviour. Our findings support the claim that $\Phi$, despite having some
theoretical drawbacks \cite{Griffith2014}, is a valuable tool for understanding
complex spatial and temporal behaviour in dynamical systems, particularly when
combined with other analysis techniques.


%


\begin{thebibliography}{17}%
\makeatletter
\providecommand \@ifxundefined [1]{%
 \@ifx{#1\undefined}
}%
\providecommand \@ifnum [1]{%
 \ifnum #1\expandafter \@firstoftwo
 \else \expandafter \@secondoftwo
 \fi
}%
\providecommand \@ifx [1]{%
 \ifx #1\expandafter \@firstoftwo
 \else \expandafter \@secondoftwo
 \fi
}%
\providecommand \natexlab [1]{#1}%
\providecommand \enquote  [1]{``#1''}%
\providecommand \bibnamefont  [1]{#1}%
\providecommand \bibfnamefont [1]{#1}%
\providecommand \citenamefont [1]{#1}%
\providecommand \href@noop [0]{\@secondoftwo}%
\providecommand \href [0]{\begingroup \@sanitize@url \@href}%
\providecommand \@href[1]{\@@startlink{#1}\@@href}%
\providecommand \@@href[1]{\endgroup#1\@@endlink}%
\providecommand \@sanitize@url [0]{\catcode `\\12\catcode `\$12\catcode
  `\&12\catcode `\#12\catcode `\^12\catcode `\_12\catcode `\%12\relax}%
\providecommand \@@startlink[1]{}%
\providecommand \@@endlink[0]{}%
\providecommand \url  [0]{\begingroup\@sanitize@url \@url }%
\providecommand \@url [1]{\endgroup\@href {#1}{\urlprefix }}%
\providecommand \urlprefix  [0]{URL }%
\providecommand \Eprint [0]{\href }%
\providecommand \doibase [0]{http://dx.doi.org/}%
\providecommand \selectlanguage [0]{\@gobble}%
\providecommand \bibinfo  [0]{\@secondoftwo}%
\providecommand \bibfield  [0]{\@secondoftwo}%
\providecommand \translation [1]{[#1]}%
\providecommand \BibitemOpen [0]{}%
\providecommand \bibitemStop [0]{}%
\providecommand \bibitemNoStop [0]{.\EOS\space}%
\providecommand \EOS [0]{\spacefactor3000\relax}%
\providecommand \BibitemShut  [1]{\csname bibitem#1\endcsname}%
\let\auto@bib@innerbib\@empty
\bibitem [{\citenamefont {Pikovsky}\ \emph {et~al.}(2001)\citenamefont
  {Pikovsky}, \citenamefont {Rosenblum},\ and\ \citenamefont
  {Kurths}}]{Pikovsky2001}%
  \BibitemOpen
  \bibfield  {author} {\bibinfo {author} {\bibfnamefont {A.}~\bibnamefont
  {Pikovsky}}, \bibinfo {author} {\bibfnamefont {M.}~\bibnamefont {Rosenblum}},
  \ and\ \bibinfo {author} {\bibfnamefont {J.}~\bibnamefont {Kurths}},\ }\href
  {http://www.cambridge.org/gb/academic/subjects/physics/nonlinear-science-and-fluid-dynamics/synchronization-universal-concept-nonlinear-sciences}
  {\emph {\bibinfo {title} {{Synchronization: A Universal Concept in Nonlinear
  Sciences}}}}\ (\bibinfo  {publisher} {Cambridge University Press},\ \bibinfo
  {address} {Cambridge},\ \bibinfo {year} {2001})\ p.\ \bibinfo {pages}
  {432}\BibitemShut {NoStop}%
\bibitem [{\citenamefont {Kuramoto}(1984)}]{Kuramoto1984}%
  \BibitemOpen
  \bibfield  {author} {\bibinfo {author} {\bibfnamefont {Y.}~\bibnamefont
  {Kuramoto}},\ }\href@noop {} {\emph {\bibinfo {title} {{Chemical
  Oscillations, Waves and Turbulence}}}}\ (\bibinfo  {publisher} {Dover
  Publications},\ \bibinfo {year} {1984})\ p.\ \bibinfo {pages}
  {164}\BibitemShut {NoStop}%
\bibitem [{\citenamefont {Panaggio}\ and\ \citenamefont
  {Abrams}(2015)}]{Panaggio2015}%
  \BibitemOpen
  \bibfield  {author} {\bibinfo {author} {\bibfnamefont {M.~J.}\ \bibnamefont
  {Panaggio}}\ and\ \bibinfo {author} {\bibfnamefont {D.~M.}\ \bibnamefont
  {Abrams}},\ }\href {\doibase 10.1088/0951-7715/28/3/R67} {\bibfield
  {journal} {\bibinfo  {journal} {Nonlinearity}\ }\textbf {\bibinfo {volume}
  {28}},\ \bibinfo {pages} {R67} (\bibinfo {year} {2015})},\ \Eprint
  {http://arxiv.org/abs/1403.6204} {arXiv:1403.6204} \BibitemShut {NoStop}%
\bibitem [{\citenamefont {Shanahan}(2010)}]{Shanahan2010}%
  \BibitemOpen
  \bibfield  {author} {\bibinfo {author} {\bibfnamefont {M.}~\bibnamefont
  {Shanahan}},\ }\href {\doibase 10.1063/1.3305451} {\bibfield  {journal}
  {\bibinfo  {journal} {Chaos}\ }\textbf {\bibinfo {volume} {20}},\ \bibinfo
  {pages} {013108} (\bibinfo {year} {2010})},\ \Eprint
  {http://arxiv.org/abs/0908.3881} {arXiv:0908.3881} \BibitemShut {NoStop}%
\bibitem [{\citenamefont {Cabral}\ \emph {et~al.}(2011)\citenamefont {Cabral},
  \citenamefont {Hugues}, \citenamefont {Sporns},\ and\ \citenamefont
  {Deco}}]{Cabral2011}%
  \BibitemOpen
  \bibfield  {author} {\bibinfo {author} {\bibfnamefont {J.}~\bibnamefont
  {Cabral}}, \bibinfo {author} {\bibfnamefont {E.}~\bibnamefont {Hugues}},
  \bibinfo {author} {\bibfnamefont {O.}~\bibnamefont {Sporns}}, \ and\ \bibinfo
  {author} {\bibfnamefont {G.}~\bibnamefont {Deco}},\ }\href {\doibase
  10.1016/j.neuroimage.2011.04.010} {\bibfield  {journal} {\bibinfo  {journal}
  {NeuroImage}\ }\textbf {\bibinfo {volume} {57}},\ \bibinfo {pages} {130}
  (\bibinfo {year} {2011})}\BibitemShut {NoStop}%
\bibitem [{\citenamefont {Hellyer}\ \emph {et~al.}(2015)\citenamefont
  {Hellyer}, \citenamefont {Scott}, \citenamefont {Shanahan}, \citenamefont
  {Sharp},\ and\ \citenamefont {Leech}}]{Hellyer2015}%
  \BibitemOpen
  \bibfield  {author} {\bibinfo {author} {\bibfnamefont {P.~J.}\ \bibnamefont
  {Hellyer}}, \bibinfo {author} {\bibfnamefont {G.}~\bibnamefont {Scott}},
  \bibinfo {author} {\bibfnamefont {M.}~\bibnamefont {Shanahan}}, \bibinfo
  {author} {\bibfnamefont {D.~J.}\ \bibnamefont {Sharp}}, \ and\ \bibinfo
  {author} {\bibfnamefont {R.}~\bibnamefont {Leech}},\ }\href {\doibase
  10.1523/JNEUROSCI.4648-14.2015} {\bibfield  {journal} {\bibinfo  {journal}
  {The Journal of Neuroscience}\ }\textbf {\bibinfo {volume} {35}},\ \bibinfo
  {pages} {9050} (\bibinfo {year} {2015})}\BibitemShut {NoStop}%
\bibitem [{\citenamefont {Schartner}\ \emph {et~al.}(2015)\citenamefont
  {Schartner}, \citenamefont {Seth}, \citenamefont {Noirhomme}, \citenamefont
  {Boly}, \citenamefont {Bruno}, \citenamefont {Laureys},\ and\ \citenamefont
  {Barrett}}]{Schartner2015}%
  \BibitemOpen
  \bibfield  {author} {\bibinfo {author} {\bibfnamefont {M.}~\bibnamefont
  {Schartner}}, \bibinfo {author} {\bibfnamefont {A.}~\bibnamefont {Seth}},
  \bibinfo {author} {\bibfnamefont {Q.}~\bibnamefont {Noirhomme}}, \bibinfo
  {author} {\bibfnamefont {M.}~\bibnamefont {Boly}}, \bibinfo {author}
  {\bibfnamefont {M.-A.}\ \bibnamefont {Bruno}}, \bibinfo {author}
  {\bibfnamefont {S.}~\bibnamefont {Laureys}}, \ and\ \bibinfo {author}
  {\bibfnamefont {A.~B.}\ \bibnamefont {Barrett}},\ }\href {\doibase
  10.1371/journal.pone.0133532} {\bibfield  {journal} {\bibinfo  {journal}
  {PloS One}\ }\textbf {\bibinfo {volume} {10}},\ \bibinfo {pages} {e0133532}
  (\bibinfo {year} {2015})}\BibitemShut {NoStop}%
\bibitem [{\citenamefont {Shanahan}(2008)}]{Shanahan2008}%
  \BibitemOpen
  \bibfield  {author} {\bibinfo {author} {\bibfnamefont {M.}~\bibnamefont
  {Shanahan}},\ }\href {\doibase 10.1103/PhysRevE.78.041924} {\bibfield
  {journal} {\bibinfo  {journal} {Physical Review E}\ }\textbf {\bibinfo
  {volume} {78}},\ \bibinfo {pages} {041924} (\bibinfo {year}
  {2008})}\BibitemShut {NoStop}%
\bibitem [{\citenamefont {Balduzzi}\ and\ \citenamefont
  {Tononi}(2008)}]{Balduzzi2008a}%
  \BibitemOpen
  \bibfield  {author} {\bibinfo {author} {\bibfnamefont {D.}~\bibnamefont
  {Balduzzi}}\ and\ \bibinfo {author} {\bibfnamefont {G.}~\bibnamefont
  {Tononi}},\ }\href {\doibase 10.1371/journal.pcbi.1000091} {\bibfield
  {journal} {\bibinfo  {journal} {PLoS Computational Biology}\ }\textbf
  {\bibinfo {volume} {4}},\ \bibinfo {pages} {e1000091} (\bibinfo {year}
  {2008})}\BibitemShut {NoStop}%
\bibitem [{\citenamefont {Barrett}\ and\ \citenamefont
  {Seth}(2011)}]{Barrett2011}%
  \BibitemOpen
  \bibfield  {author} {\bibinfo {author} {\bibfnamefont {A.~B.}\ \bibnamefont
  {Barrett}}\ and\ \bibinfo {author} {\bibfnamefont {A.~K.}\ \bibnamefont
  {Seth}},\ }\href {\doibase 10.1371/journal.pcbi.1001052} {\bibfield
  {journal} {\bibinfo  {journal} {PLoS Computational Biology}\ }\textbf
  {\bibinfo {volume} {7}},\ \bibinfo {pages} {e1001052} (\bibinfo {year}
  {2011})}\BibitemShut {NoStop}%
\bibitem [{\citenamefont {Griffith}(2014)}]{Griffith2014}%
  \BibitemOpen
  \bibfield  {author} {\bibinfo {author} {\bibfnamefont {V.}~\bibnamefont
  {Griffith}}}\href {http://arxiv.org/abs/1401.0978} ,\  \Eprint {http://arxiv.org/abs/1401.0978} {arXiv:1401.0978}
  \BibitemShut {NoStop}%
\bibitem [{\citenamefont {Tegmark}(2016)}]{Tegmark2016}%
  \BibitemOpen
  \bibfield  {author} {\bibinfo {author} {\bibfnamefont {M.}~\bibnamefont
  {Tegmark}}}\href {http://arxiv.org/abs/1601.02626} ,\ \Eprint {http://arxiv.org/abs/1601.02626} {arXiv:1601.02626}
  \BibitemShut {NoStop}%
\bibitem [{\citenamefont {Abrams}\ \emph {et~al.}(2008)\citenamefont {Abrams},
  \citenamefont {Mirollo}, \citenamefont {Strogatz},\ and\ \citenamefont
  {Wiley}}]{Abrams2008}%
  \BibitemOpen
  \bibfield  {author} {\bibinfo {author} {\bibfnamefont {D.~M.}\ \bibnamefont
  {Abrams}}, \bibinfo {author} {\bibfnamefont {R.}~\bibnamefont {Mirollo}},
  \bibinfo {author} {\bibfnamefont {S.~H.}\ \bibnamefont {Strogatz}}, \ and\
  \bibinfo {author} {\bibfnamefont {D.~A.}\ \bibnamefont {Wiley}},\ }\href
  {\doibase 10.1103/PhysRevLett.101.084103} {\bibfield  {journal} {\bibinfo
  {journal} {Physical Review Letters}\ }\textbf {\bibinfo {volume} {101}},\
  \bibinfo {pages} {084103} (\bibinfo {year} {2008})}\BibitemShut {NoStop}%
\bibitem [{\citenamefont {King}\ \emph {et~al.}(2013)\citenamefont {King},
  \citenamefont {Sitt}, \citenamefont {Faugeras}, \citenamefont {Rohaut},
  \citenamefont {{El Karoui}}, \citenamefont {Cohen}, \citenamefont
  {Naccache},\ and\ \citenamefont {Dehaene}}]{King2013}%
  \BibitemOpen
  \bibfield  {author} {\bibinfo {author} {\bibfnamefont {J.-R.}\ \bibnamefont
  {King}}, \bibinfo {author} {\bibfnamefont {J.~D.}\ \bibnamefont {Sitt}},
  \bibinfo {author} {\bibfnamefont {F.}~\bibnamefont {Faugeras}}, \bibinfo
  {author} {\bibfnamefont {B.}~\bibnamefont {Rohaut}}, \bibinfo {author}
  {\bibfnamefont {I.}~\bibnamefont {{El Karoui}}}, \bibinfo {author}
  {\bibfnamefont {L.}~\bibnamefont {Cohen}}, \bibinfo {author} {\bibfnamefont
  {L.}~\bibnamefont {Naccache}}, \ and\ \bibinfo {author} {\bibfnamefont
  {S.}~\bibnamefont {Dehaene}},\ }\href {\doibase 10.1016/j.cub.2013.07.075}
  {\bibfield  {journal} {\bibinfo  {journal} {Current Biology}\ }\textbf
  {\bibinfo {volume} {23}},\ \bibinfo {pages} {1914} (\bibinfo {year}
  {2013})}\BibitemShut {NoStop}%
\bibitem [{\citenamefont {Oizumi}\ \emph {et~al.}(2014)\citenamefont {Oizumi},
  \citenamefont {Albantakis},\ and\ \citenamefont {Tononi}}]{Oizumi2014}%
  \BibitemOpen
  \bibfield  {author} {\bibinfo {author} {\bibfnamefont {M.}~\bibnamefont
  {Oizumi}}, \bibinfo {author} {\bibfnamefont {L.}~\bibnamefont {Albantakis}},
  \ and\ \bibinfo {author} {\bibfnamefont {G.}~\bibnamefont {Tononi}},\ }\href
  {\doibase 10.1371/journal.pcbi.1003588} {\bibfield  {journal} {\bibinfo
  {journal} {PLoS Computational Biology}\ }\textbf {\bibinfo {volume} {10}},\
  \bibinfo {pages} {e1003588} (\bibinfo {year} {2014})}\BibitemShut {NoStop}%
\bibitem [{\citenamefont {Tononi}(2012)}]{Tononi2012}%
  \BibitemOpen
  \bibfield  {author} {\bibinfo {author} {\bibfnamefont {G.}~\bibnamefont
  {Tononi}},\ }\href {http://www.ncbi.nlm.nih.gov/pubmed/23165867} {\bibfield
  {journal} {\bibinfo  {journal} {Archives Italiennes de Biologie}\ }\textbf
  {\bibinfo {volume} {150}},\ \bibinfo {pages} {56} (\bibinfo {year}
  {2012})}\BibitemShut {NoStop}%
\bibitem [{\citenamefont {Barrett}(2014)}]{Barrett2014}%
  \BibitemOpen
  \bibfield  {author} {\bibinfo {author} {\bibfnamefont {A.~B.}\ \bibnamefont
  {Barrett}}}\href {http://arxiv.org/abs/1411.2832} ,\ \Eprint {http://arxiv.org/abs/1411.2832} {arXiv:1411.2832}
  \BibitemShut {NoStop}%
\end{thebibliography}
\end{document}